\documentclass[journal=jacsat,manuscript=article]{achemso}

\usepackage[version=3]{mhchem} % Formula subscripts using \ce{}
\usepackage{amsthm,amsmath,amssymb}
\usepackage{mathrsfs}
\usepackage{caption}
\usepackage{subcaption}
\usepackage{hyperref}
\usepackage[dvipsnames]{xcolor}
\usepackage{soul}

\author{Yue Jian}
\affiliation[Second University]
{Department of Materials Science and Engineering, Carnegie Mellon University, USA}

\author{Yuyang Wang}
\affiliation[First University]
{Department of Mechanical Engineering, Carnegie Mellon University, USA}
\alsoaffiliation[Third University]
{Machine Learning Department, Carnegie Mellon University, USA}

\author{Amir Barati Farimani}
\email{barati@cmu.edu}
\affiliation[First University]
{Department of Mechanical Engineering, Carnegie Mellon University, USA}
\alsoaffiliation[Second University]
{Department of Materials Science and Engineering, Carnegie Mellon University, USA}
\alsoaffiliation[Third University]
{Machine Learning Department, Carnegie Mellon University, USA}

\title[An \textsf{achemso} demo]
  { Predicting $CO_2$ Absorption in Ionic Liquids with Molecular Descriptors and Explainable Graph Neural Networks}
\abbreviations{IR,NMR,UV}

\begin{document}
\begin{sloppypar}
% \begin{tocentry}

% Some journals require a graphical entry for the Table of Contents.
% This should be laid out ``print ready'' so that the sizing of the

% Inside the \texttt{tocentry} environment, the font used is Helvetica
% 8\,pt, as required by \emph{Journal of the American Chemical

% The surrounding frame is 9\,cm by 3.5\,cm, which is the maximum
% permitted for  \emph{Journal of the American Chemical Society}
% graphical table of content entries. The box will not resize if the
% content is too big: instead it will overflow the edge of the box.

% This box and the associated title will always be printed on a
% separate page at the end of the document.

% \end{tocentry}

\begin{abstract}
Ionic Liquids (ILs) provide a promising solution for $CO_2$ capture and storage to mitigate global warming. However, identifying and designing the high-capacity IL from the giant chemical space requires expensive, and exhaustive simulations and experiments. Machine learning (ML) can accelerate the process of searching for desirable ionic molecules through accurate and efficient property predictions in a data-driven manner. But existing descriptors and ML models for the ionic molecule suffer from the inefficient adaptation of molecular graph structure. Besides, few works have investigated the explainability of ML models to help understand the learned features that can guide the design of efficient ionic molecules. In this work, we develop both fingerprint-based ML models and Graph Neural Networks (GNNs) to predict the $CO_2$ absorption in ILs. Fingerprint works on graph structure at the feature extraction stage, while GNNs directly handle molecule structure in both the feature extraction and model prediction stage. We show that our method outperforms previous ML models by reaching a high accuracy (MAE of 0.0137, $R^2$ of 0.9884). Furthermore, we take the advantage of GNNs feature representation and develop a substructure-based explanation method that provides insight into how each chemical fragments within IL molecules contribute to the $CO_2$ absorption prediction of ML models. We also show that our result agrees with some ground truth on functional group importance from the theoretical understanding of $CO_2$ absorption in ILs, which can advise on the design of novel and efficient functional ILs in the future.
\end{abstract}

\section{Introduction}
Global warming is a major environmental problem in our world. Based on the prediction of the Intergovernmental Panel on Climate Change (IPCC), the average temperature of our world will rise about $1.9^\circ C$ if we don't take any action by 2100\cite{Stewart2005}. Among all of the greenhouse gases, $CO_2$ makes the most contribution to global warming to an extent of about $78.6\%$\cite{mis}. How to effectively capture and store $CO_2$ is crucial for solving the global warming problem. Existing methods, including physisorption/chemisorption\cite{Mandal2005,Barzagli2009}, membrane separation\cite{Ebner2009} or molecular sieves\cite{Zelenak2008}, carbamation, amine physical absorption\cite{Chaffee2007}, amine dry scrubbing\cite{Serna-Guerrero2008}, and mineral carbonation\cite{Liu2005,Druckenmiller2005}, have been introduced to absorb $CO_2$. However, the reagents used in these methods suffer from insufficient carbon dioxide storage capacity, high energy demand in absorption process and low thermal stability.\cite{Hu2018,Zhang2020} The evaporation and degradation of reagents may lead the storage process to become costly \cite{Dawodu1996}. 

Ionic liquids (ILs) are families of molten salt that remains liquid state at room temperature. Over the past decades, it received significant attention and has been an intensive research area due to its unique physical and chemical properties, such as nonvolatility, high chemical stability, high $CO_2$ solubility, and easy operation at liquid state. Those properties make ILs an ideal candidate for $CO_2$ storage \cite{Lei2014,Domanska2011,Seiler2004,Lei2008,Diedenhofen2010,Dong2007,Marsh2004,Meyer2003,Welton1999}. Usually, IL composes pair of ions with different charges, and the combination of ions largely determine the properties of ILs. However, such combinations of cations and anions as well as the various selections of cations and anions themselves make it challenging to exhaust the design space of IL for efficient $CO_2$ storage through experiments. To efficiently estimate the $CO_2$ absorption of ILs, researchers have investigated the quantitative structure-property relationship (QSPR). QSPR method aims at building mathematical models for the prediction of numerical properties based on structural information of chemical compounds \cite{NarayanDas2013,Cherkasov2014}. But traditional methods used in QSPR such as Molecular Dynamic (MD) and Density Function Theory (DFT) can be computationally challenging for ILs due to the complexity of inter- and intra- molecular interaction\cite{Koutsoukos2021,Abdulfatai2019,Puzyn2008}.

The recent development of Machine Learning (ML) methods bears the promise for QSPR modeling through accurate and efficient property predictions of chemical compounds. Compared with conventional simulation methods like MD or DFT\cite{Li2022}, ML methods have demonstrated similar accuracy but with less computational cost in various chemical applications.\cite{Faber2017,Lawler2021,Xu2022} Especially, several works have explored applying ML models to solving ionic liquid problems via various descriptors of IL molecules. Group Contribution Theory (GC) is one of the earliest descriptors for IL molecules \cite{Valderrama2009,Paduszynski2014,Gardas2008,Matsuda2007}. GC manually breaks down the molecule into different characteristic functional groups and counts the existence frequency for each group. However, this descriptor is highly human experience-dependent and may lead to the loss of information for substructures within the group. Another way of finding molecule descriptors is Quantum Chemical Descriptor (QC), which utilizes the calculated properties from DFT to provide sub-molecule level representations for IL molecules\cite{Venkatraman2017,Eike2003,Mehrkesh2016}. But to gain QC descriptors, one needs to perform expensive and time-consuming QC calculations like DFT to acquire the properties. ML models like Support Vector Machine (SVM), Random Forest (RF), and deep learning models such as Multi-layer Perceptron (MLP), Convolutional Neural Network (CNN), and Recurrent Neural Network (RNN), have been applied on top of the descriptors to perform various properties prediction tasks\cite{Venkatraman2018,Deng2019}. However, both GC and QC descriptors can lack the modeling of the structural information of molecules, which confines the performance of ML models. Other molecular descriptors like Extended-Connectivity Fingerprints (ECFPs), create a feature vector by iteratively aggregating the neighbor information of each atom and hashing that into a vector \cite{Morgan1965,Rogers2010}. Such methods directly better encode the structural information of molecules and can be more expressive. However, such molecular fingerprints (FP) and how different ML models built upon them perform have not been well studied for $CO_2$ absorption in ILs. 

Recently, GNNs has shown to be a powerful tool for molecule features representation, and properties prediction and have received a significant amount of attention\cite{Atz2021,Wang2022,Wang2022a,Kipf2016,Magar2022,Yadav2022,karamad2020orbital}. At the feature representation stage, GNNs directly work on the molecular structure. It treats the molecule as a graph and utilizes an adjacent matrix to encode the bond edge and connectivity, as well as a node feature matrix to encode the atom and related properties. This representation is more generalizable, stable, and less computationally expensive compared with GC and QC descriptors.  At the model training and prediction stage, GNN aggregates the node message through edge during the forward process,\cite{Kipf2016,Brody2021,Xu2018} and it outperforms other Neural-Network-based models on unstructured graphical data\cite{Kipf2016}. GNNs have been involved in many areas related to molecules, such as drug discovery, quantum chemistry, and structural biology\cite{Gilmer2017}. But existing works using GNNs on ILs tend to focus solely on one family of anions or cations, which still needs to be expanded and generalized\cite{Ruza2020,Wang2020}.

Besides building ML models to obtain accurate and efficient predictions, how to explain the output from ML models given certain input data is also an active research area\cite{Koutsoukos2021,Jimenez-Luna2020}. In traditional experimental and computational chemistry, researchers heavily rely on their knowledge and experience in designing new compounds. On the other hand, as a black box, the intermediate decision process of the ML models is hard to unveil. Understanding how ML models make decisions can provide us with extra insights into how the structure of the input molecule affects the property of IL and new IL design from a data-driven perspective. Explainable algorithms have been developed on GNNs to analyze the importance of each input edge\cite{Ying2019}. But in the ILs research area, researchers usually focus on the prediction power of GNN on various properties but ignore the explainability of the GNN model. Benefiting from graph representation, GNNs has the potential to provide an explanation of the molecular structure importance that reaches atom and bond levels.

% fingerprint, GNNs (single graph), Explanation ( global node, atom importance)
In this paper, we introduce two categories of methods for $CO_2$ solubility prediction, namely, FP-based machine learning models and GNNs. Besides, we also developed an explanation method for IL molecule substructure importance analysis. For FP-based machine learning models, GC and FP are included as descriptors. We then compare the expressiveness of FP with GC on different machine learning models. For GNNs part, we included Graph Convolutional Networks (GCN), Graph Attention Networks (GAT), and Graph Isomorphism Networks (GIN) to do the $CO_2$ solubility prediction. Moreover, we make two improvements to data representation and the GNN framework in order to build an IL explainer. Firstly, instead of treating cation and anion as two separate graphs, we treat them as a single unconnected graph and feed the whole graph into one GNN network. Secondly, we substitute the final pooling layer of GNN with a global node that connects with every atom within one data point. Based on that, we develop an IL molecule explainer by combining the improved GNN framework with the sub-graph-based GNN explaining methods\cite{Ying2019}. Benefiting from the two improvements we make, the IL molecule explainer can provide an importance score insight into a single atom level within the IL molecule. we also find that our explanation method can provide a reasonable fragments importance ranking for the IL molecule in the prediction task and can be a useful tool to guide the design of new IL molecules. To the best of our knowledge, this is one of the first works in applying GNNs and the fragments importance explanation study that reaches the single atom level for $CO_2$ absorption in ILs.

% In this paper, we introduce two descriptors that directly work on IL molecule structures, i.e., fingerprint and graph representation. We then combine the descriptors with various ML models including Support Vector Machine (SVM), RF (RF), XGBoost, Multi-layer Perceptron (MLP) as well as Graph Convolution Network (GCN), Graph Attention Network (GAT), Graph Isomorphism Network (GIN) to perform prediction task for $CO_2$ solubility in IL. Furthermore, we developed an explanation method based on the GNN models, which can provide a bond, atom, and fragment importance explanation for IL structure with a fine-tuned GNN model. We show that the Mean Absolute Error (MAE) for fingerprint and graph representation reaches as low as 0.0151 and 0.0137, respectively, and both outperform previous ML methods. Besides, we also find that our explanation method can provide a reasonable fragments importance ranking for the IL molecule in the prediction task and can be a useful tool to guide the design of new IL molecules. To the best of our knowledge, this is a pioneering work in applying GNNs and the fragments importance explanation study that reaches the single atom or bond level for $CO_2$ absorption in ILs.

\section{Method}

\begin{figure}
  \centering
  \includegraphics[width = 15cm]{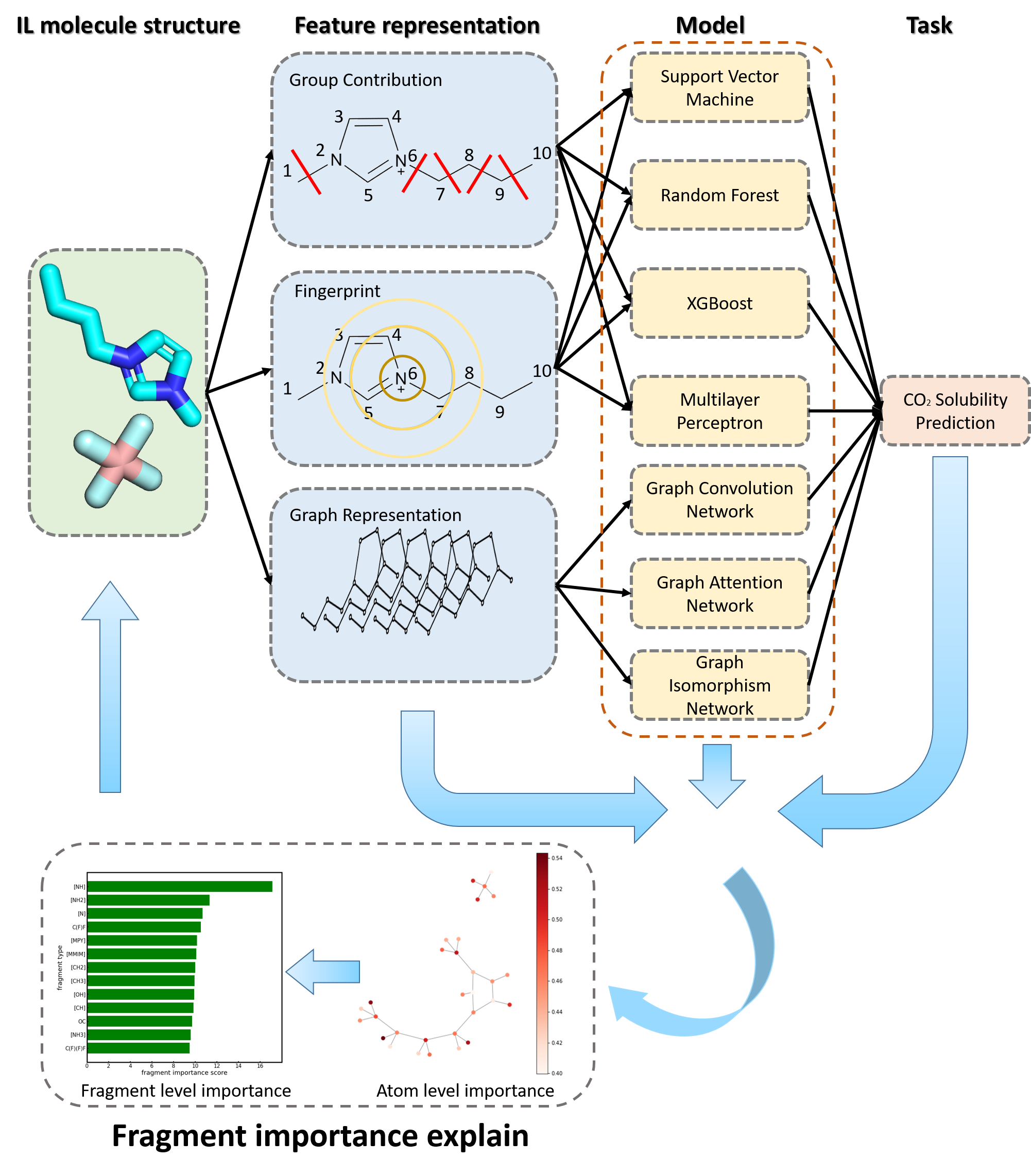}
  \caption{An overview of methods in the paper. We first use FP as a descriptor and compare it with GC. Based on the descriptors, Shallow Machine Learning models (SVM, RF, XGBoost), Deep Learning models (MLP), are built on top of the descriptors for $CO_2$ solubility prediction. Besides descriptor-based machine learning, GNN-based models (GCN, GAT, GIN) are employed to perform the solubility prediction. Furthermore, we develop an explanation method (IL explainer) for ionic molecules that can take in a fine-tuned GNN model, and an IL data point, and return the fragment importance of the molecule. We finally make the importance explanation for the functional groups within cations across the whole dataset.}
  \label{overview}
\end{figure}

Fig.~\ref{overview} is the overview of the whole work. IL molecule pairs are represented in three ways which are GC, FP, and graph representation. GC and FP are combined with various Machine Learning models such as SVM, RF, XGBoost, and MLP to perform solubility prediction tasks. For GNNs, we utilize three GNN frameworks which are GCN, GAT, and GIN for property prediction. Furthermore, an explanation method is developed and can provide both atom level and fragment level importance analysis for IL molecule pair by taking in a fine-tuned GNN model and a target IL molecule pair.

% In this work, we included both descriptor-based machine learning methods and GNN-based methods for $CO_2$ solubility prediction. Specifically, the descriptor-based machine learning method utilizes Group Contribution (GC) (set as the control group) and Fingerprint (FP) as descriptors. Based on the descriptor, we build various machine learning models to perform the $CO_2$ solubility prediction task. For the GNN-based method, we employed different message-passing models to predict $CO_2$ solubility directly from graphical structures of anion and cation molecules. Besides, we also develop a model-explaining method for analyzing fragment importance within the molecule. Furthermore, the explaining method can provide fragment importance by feeding in the trained GNN model and input data. The overview of the whole framework is shown in Fig.~\ref{fgr:example}.

\subsection{IL Dataset}
The dataset contains 10,117 data points of $CO_2$ solubility in various ionic liquids (ILs) systems at different temperatures and pressure\cite{Song2020}, which is initially collected and published by Lei $et \ al$\cite{Lei2014}. Each data point includes the SMILES of cation, anion, temperature, pressure, and the solubility of $CO_2$ (expressed as the mole fraction of $CO_2$ in ionic liquid). Solubilities of ILs range from 0.0000648 to 0.9516 in mole fraction, the temperatures range from 243.2K to 453.15K, and the pressures range from 0.00798 bar to 499.9 bar. Cations of ILs include imidazolium, pyrrolidinium, pyridinium, piperidinium, ammonium, phosphonium, and sulfonium. Anions of ILs contain tetrafluoroborate $[BF_4]$, chloride $[Cl]$, dicyanamide $[DCA]$, nitrate $[NO_3]$, hexafluorophosphate $[PF_6]$, thiocyanate $[SCN]$, tricyanomethanide $[C(CN)_3]$, hydrogen sulfate $[HSO_4]$, bis (trifluoro methylsulfonyl) amide $[Tf_2N]$, methylsulfate $[MeSO_4]$, etc. Following previous works, the dataset is randomly split into training and test sets by the ratio of 80\% and 20\%. 

\subsection{Descriptor-based Machine Learning Models}

\subsubsection{Descriptor engineering}
We utilize the Morgan fingerprint as the descriptor for the IL molecule pairs. Morgan FP is a method for generalizing molecular signatures with molecule structure information.\cite{Morgan1965} We use RDKit to generate FP.\cite{rdkit} Specifically, for the generation hyperparameters, the radius is set to be 3, and the number of bits is 2048 for each molecule. For a certain ionic molecule pair, we first obtain the FP for each molecule, then concatenate them into a vector with a length of 4096. By adding the temperature (K) and pressure (bar), the length of the final vector is 4098. We also included GC descriptor as a baseline method to compare with FP\cite{Song2020}. For GC, we slice all the ionic molecules into 51 fragments, each molecule pair are then mapped to a vector with a length of 51, and each element in the vector counts the existing frequency of each fragment in a certain molecule pair. After that, temperature and pressure are added to the vector. The length of the final vector would be 53.

% We featurize ionic liquid molecules with two representation methods: Group Contribution (GC) \cite{Song2020} and Morgan fingerprint \cite{Morgan1965}. The first one is the Group Contribution method (GC), which slices the molecule into different functional groups following the same way in the original dataset\cite{Song2020}. The dataset slices all the ionic molecules into 51 fragments, each molecule pair are then mapped to a vector with a length of 51, and each element in the vector counts the existing frequency of each fragment in a certain molecule pair. After that, we append temperature and pressure to the vector. The length of the final vector would be 53. We use GC as the control group. The second method is extracting features with Morgan fingerprint\cite{Morgan1965}. Morgan fingerprint is a method for generalizing molecular signatures with molecule structure information. In this work, we use RDKit to generate fingerprints. Specifically, for the generation hyperparameters, the radius is set to be 3, and the number of bits is 2048 for each molecule. For a certain ionic molecule pair, we first obtain the fingerprint for each molecule, then concatenate them into a vector with a length of 4096. By adding the temperature (K) and pressure (bar), the length of the final vector is 4098.

\subsubsection{Descriptor-based machine learning models}
We use 4 ML models including SVM\cite{Boser1992}, RF\cite{Ho1995}, XGBoost\cite{Chen2016}, and MLP\cite{Rumelhart1986} to comprehensively investigate the molecular descriptors. Those models take input as the descriptors as well as the conditions (i.e., temperature, pressure) and predict $CO_2$ solubility. The details and hyperparameters of the ML models can be found in the ``Model Details" section in supporting information.

\subsection{Explainable GNN}
In this section, we first introduce molecular graphs and GNNs models for IL $CO_2$ solubility prediction that is shown in Fig.~\ref{GNN}. We then conduct a post-hoc interpretability study on the GNN model by generalizing an explanation for the input molecule through the GNN model shown in Fig.~\ref{explain}. 

\subsubsection{IL Molecular Graph}
\begin{figure}

  \centering
  \includegraphics[width = 16cm]{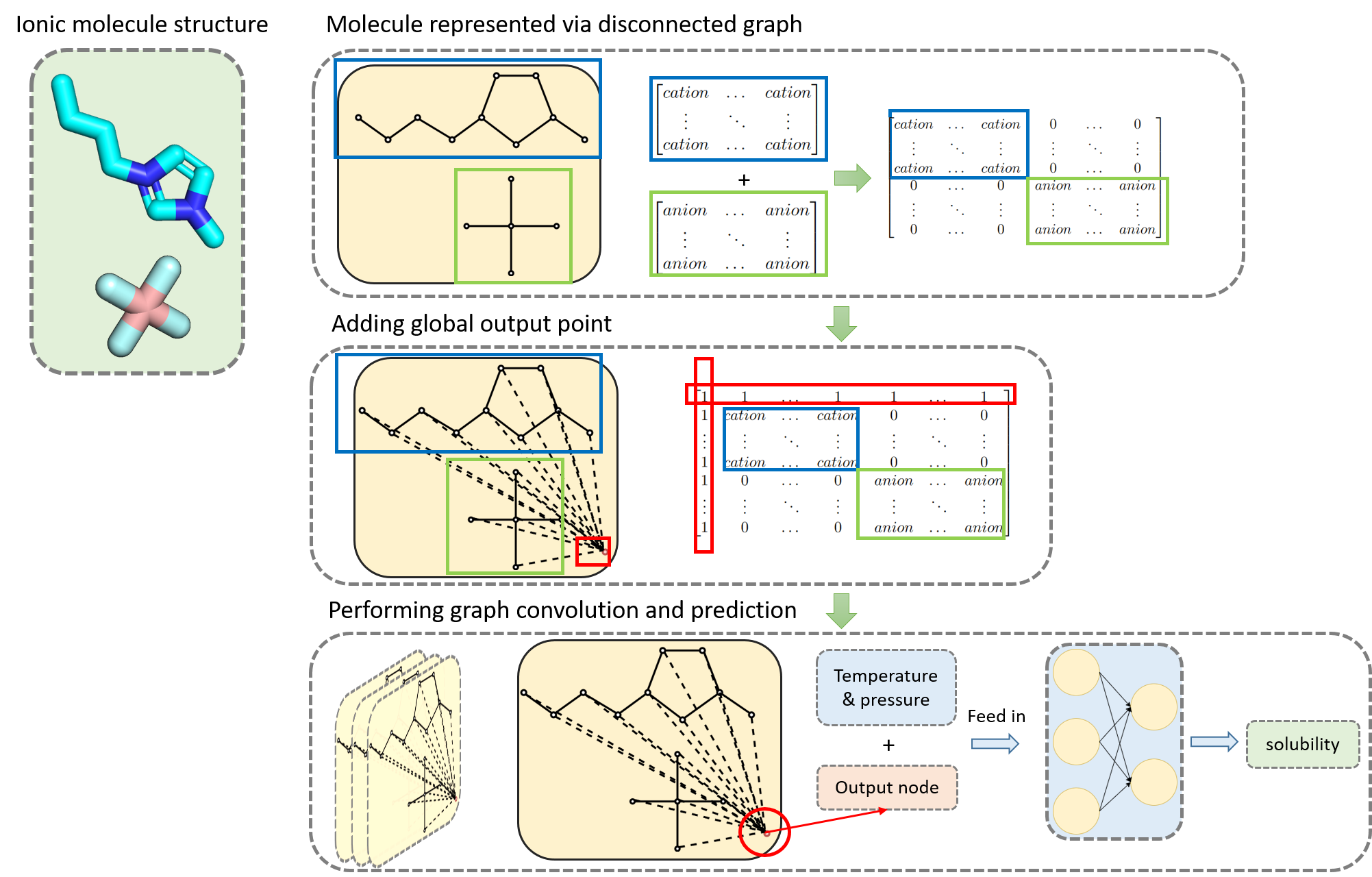}
  \caption{An illustration of how GNN predicts solubility for ionic molecule pairs. Firstly, cation and anion are treated as two separate graphs with an adjacent matrix. Then we concatenate two molecule graphs diagonally. Further, a global node is added to the graph. Finally, two molecule graphs with a global node are merged into a single undirected graph. The final graph will be fed into GNN, after several layers of message passing, the message of the global node is extracted and put into the final classification layer to get the solubility result.}
  \label{GNN}
\end{figure}
Graph representation treats a molecule as a graph $G$ with a set of nodes $V$ and edges $E$\cite{Kipf2016}. In our case, given an ionic molecule, each atom can be treated as a node while bonds can be edges. Further, a node matrix $H \in \mathbb{R}^{N\times M}$, where $N$ denotes the number of nodes and $M$ denotes the embedding dimension of node features), and an adjacent matrix $A \in \mathbb{R}^{N\times N}$ is used to represent the bond connectivity of the molecule graph. Atomic number, hybridization, aromatic, atom degree, and charge are included in atom features. Bond type, whether the bond belongs in the ring, and whether it is an aromatic bond are included in edge features. Moreover, each data point contains a cation and an anion, instead of treating both molecules separately with two graphs, we treat them as one single unconnected graph. Mathematically, we concatenate two adjacent matrices diagonally and two node matrices vertically. The way we concatenate the matrix is shown in Fig.~\ref{GNN}. Fig~\ref{GNN} also describes the whole process of using GNN to predict $CO_2$ solubility. Furthermore, to accommodate the need for IL explainer development, we substitute the final pooling layer of the model with a global node. This node takes the responsibility of outputting the final solubility and is connected to each node within the molecule graph. The details and the intuition of adding this global node will be introduced in the following section. 

\subsection{Graph Neural Network}

Graph Neural Network (GNN) is a deep learning method that directly works on graph-structured data. Let $G$ denote a graph, $V$ denote the set of nodes belonging to graph $G$, and $E$ denote a set of edges of the graph $G$. In $k$th layer of GNN, model takes a set of node representation $\{ h^{(k-1)}_v|v \in \mathcal{N}_{(v)} \}$ as input, then perform a feature aggregation as Eq.~\ref{fea_aggre} and update every node feature in the set $\{ h^{(k-1)}_v|v \in \mathcal{N}_{(v)} \}$ to $\{ h^{(k)}_v|v \in \mathcal{N}_{(v)} \}$. Here in Eq.~\ref{fea_aggre}, $\mathcal{N}_{v}$ denotes the set of neighbor nodes of node $v$, $f_\theta$ denotes an update function for all the aggregated features in the last layer\cite{Hu2019,Xu2018,Brody2021,Kipf2016,Velickovic2017}.

\begin{equation}
% general GNN node update function
h^{(k)}_{v} = f_{\theta}(h^{(k - 1)}_{v},Aggregate(\{ h^{}_{} | j \in \mathcal{N}_{(v)} \}))
\label{fea_aggre}
\end{equation}

In this work, three widely-used and powerful GNN models are applied to perform $CO_2$ solubility prediction, which develops different aggregation and combination operations to learn from molecular graphs. GCN \cite{Kipf2016} aggregates and update the node features by summing up all of the normalized node features of the neighbors for a single node and multiplying the sum with a learnable weight matrix. The update function is shown in Eq.~\ref{GCN}. 
% However, GCN can be improved in model expressiveness in two ways, firstly, the aggregation and update function of GCN is a surjection function and can lead to a loss of information in some cases during the graph convolution process.\cite{Xu2018,Hu2019} Secondly, the feature aggregation process weighs all the neighbor node feature with equal importance.\cite{Velickovic2017,Brody2021} Both parts above can limit the expressiveness of the model.

\begin{equation}
% GCNconv node update function
h^{(k)}_{v} = \mathbf{W}^\mathrm{T} \sum_{j \in \mathcal{N}_{(v)} \cup \{ v \} } (\frac{e_{j,v}}{\sqrt{\hat{d}_{j}\hat{d}_{v}}} h^{(k-1)}_{v})
\label{GCN}
\end{equation}

GIN \cite{Xu2018,Hu2019} is introduced to retain the information during the aggregation process by using an MLP as an update function as shown in Eq.~\ref{GIN}. Here $\epsilon^{(k)}$ is a learnable parameter. Different from GCN that use mean pooling as the update function, GIN substitute mean pooling with MLP. This change makes the update function from surjective to injective, and thus enhances the expressiveness of the model.
\begin{equation}
% GIN update function
h^{(k)}_{v} = \ce{MLP}^{(k)}((1 - \epsilon^{(k)})\cdot h^{( k - 1 )}_{v} + \sum_{j \in \mathcal{N}_(v)}h^{(k-1)}_j)
\label{GIN}
\end{equation}

GAT \cite{Velickovic2017,Brody2021} improve the expressiveness of GNNs by deploying the attention mechanism. Graph attention assigns a learnable weight for each edge when performing feature aggregation on nodes. In this way, the model can learn a weight on each edge during the training process and thus weigh each neighbor node with the difference in importance during the forward process of the model. The update function and the calculation of attention are given in Eq.~\ref{GAT_1} and Eq.~\ref{GAT_2}.  
\begin{equation}
% GATv2 update function
h^{(k)}_{v} = \alpha_{v,v} \mathbf{W}_{h_{v}} + \sum_{j \in \mathcal{N}_{(v)}}\alpha_{v,j}\mathbf{W}_{h,j}
\label{GAT_1}
\end{equation}

\begin{equation}
% GATv2 attention function
\alpha_{v,j} = \frac{\exp(\mathbf{a}^\mathrm{T}LeakyReLU(\mathbf{W}[h_{v}||h_{j}]))}{\sum_{k \in \mathcal{N}_{(v)} \cup {\{ v \}}}\exp(\mathbf{a}^\mathrm{T}LeakyReLU(\mathbf{W}[h_{v}||h_{k}]))}
\label{GAT_2}
\end{equation}

In this work, we include GCN, GIN, and GAT for $CO_2$ solubility prediction. All the GNN models are developed with PyTorch and PyTorch Geometric\cite{Fey/Lenssen/2019}. Details can be found in the ``GNN models" section in supporting information.

\subsubsection{IL explainer}
Subgraph-based GNN explaining method provides an explanation for the importance of edges and nodes in a graph by optimizing the mask value that is added to the adjacent matrix.\cite{Ying2019} We want to develop an explainer for the IL molecule graph based on this method so that the explainer can provide an importance score that reaches the atoms and fragments levels. Let $G$ denote a computational graph, and $X$ denote node feature information. Consider the $CO_2$ solubility prediction as a graph classification problem, then a prediction process can be seen as $\hat{y} = \Phi(G,X)$, where $\Phi$ is GNN model, $\hat{y}$ is predicted result of $CO_2$ solubility in a specific ionic molecule pair. An important question is which fragments in the IL molecule contribute most to the $CO_2$ absorption in ILs. To get this insight from trained GNNs in a data-driven manner, we want to find a subgraph $G_s$ where $G_s \subseteq G$ such that $G_s$ is important to the prediction for $\hat{y}$. Mathematically, we can formalize the above process as an optimization framework below:
\begin{equation}
\max_{G_s} MI(Y,G_s) = H(Y) - H(Y|G = G_s)
\end{equation}
Here MI denotes Mutual Information (MI) that can quantify the change in probability for the prediction $\hat{y} = \Phi(G, X)$ between the unconditional prediction and the prediction that condition on subgraph $G_s$. If we find a subset $G_s$ that maximizes the mutual information, we can say the subset has importance for the GNN prediction process. Since the entropy term $H(Y)$ is fixed as $\Phi$ is fixed for a trained GNN, our optimization process is equivalent to:
\begin{equation}
\min_{G_s} H(Y|G = G_s)
\end{equation}

However, it can be hard to traverse all the subgraphs for a whole molecular graph and compute the MI, because the number of the subgraph grows exponentially with the numbers of the nodes and edges. To solve the problem, instead of straightly masking off the part that does not belong to the subgraph, a learnable mask matrix with all of its element values between 0 and 1 is assigned to each edge in the graph. Based on that, we further optimize the mask value on each edge with gradient descent in order to maximize the mutual information. Finally, a traversal problem is reduced to an optimization problem that can be solved with gradient descent. Let $\sigma(M)$ denote the mask for graph adjacent matrix, where $M$ is the original mask $M \in \mathbb{R}^{n\times n}$, and sigmoid function $\sigma$ is to map the element in $M$ to $(0,1)$. Finally, the optimization framework can be written as:

\begin{equation}
\min_{G_s} H(Y|G_s = A \odot \sigma(M))
\label{target}
\end{equation}
Here $A$ denote the adjacent matrix for graph $G$, and the objective is to learn a mask $\sigma(M)$ such that it can minimize the target function above. In our experiment, to optimize a mask for a single molecule pair, we run 100 epochs for each pair to let the target function Eq.~\ref{target} converge and use the optimized mask for importance score computing.

\begin{figure}

  \centering
  \includegraphics[width = 16cm]{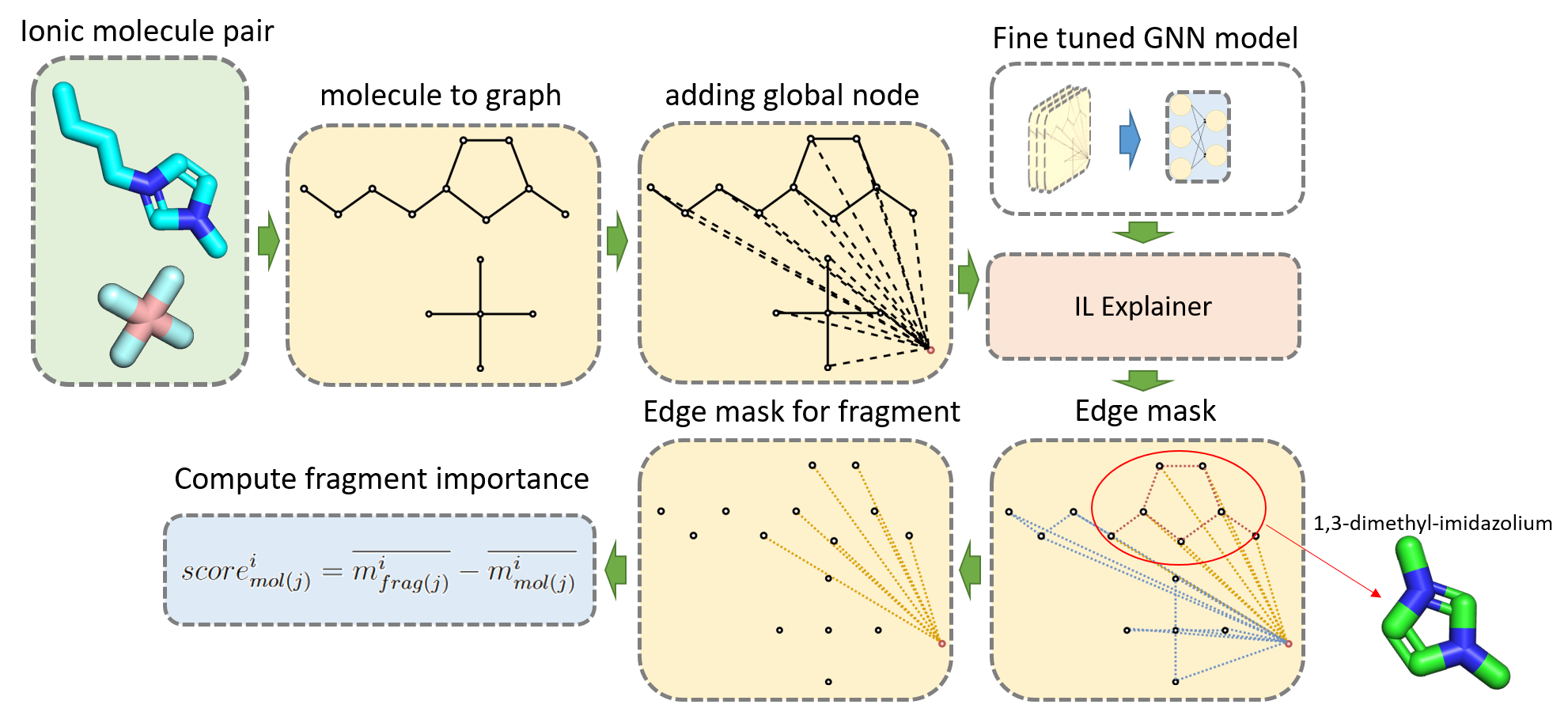}
  \caption{Making explanation on an ionic molecule pair and extracting the fragment importance of 1,3-dimethyl-imidazolium from the edge mask. To do the explanation, we still need an ionic molecule pair and process it into a graph. We feed the molecule graph and a fine-tuned GNN model (this could be GCN, GAT, GIN or etc.) into the IL explainer. Then we run the explainer to do the optimization on the mask matrix. Finally, we get an edge importance mask for the input graph. The importance of the edge between the atom node and the global node is seen as the importance of the atom. After that, we normalize and compute the fragment importance based on the importance of atoms that the fragment includes.}
  \label{explain}
\end{figure}

This method provides an explanation of the importance of each edge (bond) for the molecule graph, but could not provide that for node (atom). In that case, it is challenging to derive the importance of fragments in the $CO_2$ solubility prediction of ILs since some of the fragments may only contain one atom and do not have edges in their molecule graph. To gain fragment importance from a node perspective, we leverage the global node added to the cation and anion pair. Since the global node feature is the representation of the IL and is connected to all atoms in the molecular graph, we can consider the importance of edges that are connected between the global node and atom as the importance of each atom to the prediction result. We can then use this atom-level importance to formalize importance analysis for a specific functional group for the whole dataset further. Fig.~\ref{explain} illustrates the explanation process and how to extract fragment importance score from a single ionic molecule pair using 1,3-dimethyl-imidazolium as an example. 

Following the idea, we developed a score function for quantifying the importance of a specific functional group with respect to the whole dataset. Let $M_{mol}$ denote a set of edge masks values (importance value) that includes all edges between the global node and each atom node within the same molecule, $\overline{M_{mol}}$ denote the mean value of the set $M_{mol}$. Besides, let  $M_{frag}$ denote a set of edge masks values that only contain edges between the global node and the atom for a specific functional group fragment that we are interested in, $\overline{M_{frag}}$ denotes the mean value of the set $M_{frag}$. We compute the importance score of a single type of fragment $i$ within a single ionic molecule pair $j$ as follows:
\begin{equation}
score_{mol(j)}^i = \overline{M_{frag(j)}^i} - \overline{M_{mol(j)}^i}
\end{equation}
Further, the importance score of fragment $i$ within the whole dataset $Score^i$ is computed as:
\begin{equation}
Score^i = \frac{\sum_{j = 1}^{N_{frag}}score_{mol(j)}^i}{N_{frag}} = \frac{\sum_{j = 1}^{N_{frag}}(\overline{M_{frag(j)}^i} - \overline{M_{mol(j)}^i})}{N_{frag}}, 
\end{equation}
where $N_{frag}$ denotes the number of ionic molecule pairs within the whole dataset that contain the certain fragment $i$. By this means, the score variance is normalized between different IL pairs to an averaged importance score for each functional group. Our goal is to utilize the score function above to analyze the ranking of importance of different functional group fragments in the dataset's IL molecule space. We collected a set of functional groups that exists in the cations of the dataset and compute the importance ranking of those functional group. Then we compare the ranking with the theoretical evidence for $CO_2$ solubility in IL to test if the IL explainer can provide a reasonable explanation for the IL molecule.

\section{Results and Discussions}

\subsection{$CO_2$ solubility prediction}
 
\begin{figure}

  \centering
  \includegraphics[width = 10cm]{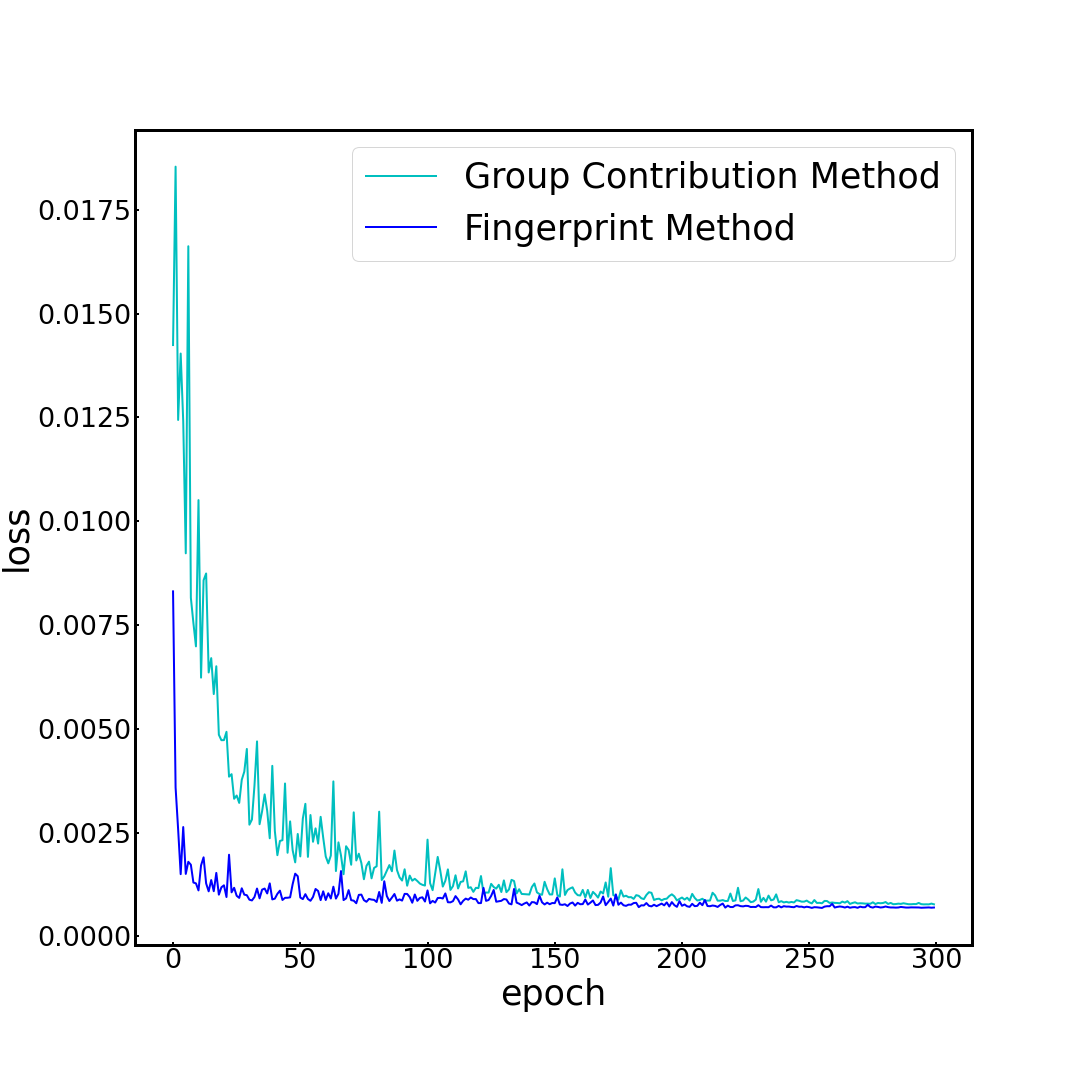}
  \caption{Validation loss-epoch curve for both GC and FP.}
  \label{converge}
\end{figure}

The result of the $CO_2$ solubility prediction tasks is shown in Table~\ref{pred_result}. Model performance is measured by MAE and $R^2$ on the test set. MAE can quantify the accuracy of the predicted result, the lower MAE is the more accurate the predicted results are. $R^2$ quantifies the proportion of the variation in the predicted solubility from the input data, and higher $R^2$ demonstrates better model performance. The result shows that FP outperforms GC on SVM, RF as well as MLP with the evaluation through MAE and $R^2$. Fig.~\ref{converge} shows the model with FP representation also converges much faster than the model with GC representation. After 300 epochs of training, FP still reaches a higher accuracy than the GC method. Though Morgan FP does not surpass GC using XGBoost, the differences between them on MAE and $R^2$ are trivial. The reason that FP generally outperforms the GC method is that the GC method can lead to information loss during the representation process since it manually sliced molecules into a fragment and ignore the structure details within a fragment. As for GNN models, GIN is the best-performing model compared with GCN and GAT. GIN also achieves the best performance among all the ML models. Firstly, except for GCN, GAT and GIN outperform most of the other descriptor-based machine learning groups. This proves the power of the GNNs model. One possible explanation for the uncompetitive performance of GCN is that although GCN uses graph representation like the other two GNN models, the feature aggregation and update process are much simpler than the other two models, also the update function of GCN is a surjective function. These will lead to a relative loss of information during the graph convolution process and finally result in a worse performance. Secondly, though FP representation and the GNN model both work on the graph structure of IL molecule, FP only handles graph structure at the feature representation stage while both feature representation and the post-feature-extraction stage for GNN are all designed to handle graph-structured data. For these reasons, a GNN with a good expressiveness (here like GIN) is likely to beat the descriptor-based method on the task related to molecule and this is also proved by the experiment result in Table~\ref{pred_result}.  

\begin{table}[t]
\centering
\caption{Performance of different Models combined with different feature representations on $CO_2$ solubility prediction task, MAE and $R^2$ for different combinations are reported here. Noted that: To make sure the model converges, we run 300 epochs to train each model below and finally take the one with the lowest validation loss as the result($\uparrow$:the higher the better,$\downarrow$: the lower the better)}
\begin{tabular}[t]{lcc}
\hline
Model  & MAE$\downarrow$& $R^2 \uparrow$\\
\hline
SVM + GC & 0.0753 & 0.8240\\
SVM + FP & 0.0655 & 0.8633\\
RF + GC & 0.0223 & 0.9774\\
RF + FP & 0.0209 & 0.9802\\
XGBoost + GC & 0.0182 & 0.9865\\
XGBoost + FP & 0.0189 & 0.9847\\
MLP + GC & 0.0170 & 0.9873\\
MLP + FP & 0.0151 & 0.9883\\
GCN & 0.0723 & 0.8197\\
GAT & 0.0253 & 0.9767\\
GIN & \textbf{0.0137} & \textbf{0.9884}\\
\hline
\end{tabular}
\label{pred_result}
\end{table}

\subsection{Model Explanation Result Analysis}

\subsubsection{Theoretical understanding of $CO_2$ solubility in ILs}
Before diving into the explanation of the machine learning model in a data-driven manner, it is necessary to check the theories about the $CO_2$ absorption process in ILs. Such theories validate the explanation results from our IL explainer. However, there are two major challenges in validating the data-driven explanation results. Firstly, existing ground truth with respect to some of the fragments distributed sparsely in various literature. Secondly, the factor that influences the $CO_2$ in solubility in a certain IL system is complex. One fragment can influence the solubility in a physical way or chemical way to different extents, which means the importance of one fragment can compose of various mechanisms. In the following section, we summarize and categorize the theoretical explanation in the literature.

The mechanism of $CO_2$ solvation in IL can be divided into physical and chemical absorptions depending on the interaction between $CO_2$ and the ILs. Physical absorption considers anions that play the primary role in $CO_2$ storage\cite{Anthony2005}. However, according to the experiment record, when the anion is unchanged, the fluorination of alkyl chain and ester groups on cations is favorable for improving $CO_2$ solubility\cite{Muldoon2007,Switzer2014,Zeng2017}. Another model based on physical absorption is called the free volume model. This model suggests that ionic molecules especially cations with longer alkyl chains tend to create more free volume space for trapping $CO_2$ inside the liquid. Thus carbon alkyl chain is favorable for the $CO_2$ storage process as well. Chemical absorption which depends on chemical interaction between $CO_2$ and IL molecules, on the other hand, usually results in a stronger combination between $CO_2$ and IL than physical absorption.  This is because the chemical bond is usually stronger than physical bond. Generally, the chemical bond is mainly composed of Coulomb interaction while the major contribution of physical interaction is Van der Waals force\cite{Zeng2017}. One of the major chemical interactions for $CO_2$ and IL is between Amine and $CO_2$. $CO_2$ tends to react with the amine in IL in a similar manner as aqueous amine by forming a carbamate salt. Amine-$CO_2$ interaction is one of most strengthen interaction between IL and $CO_2$. There are also Non Amine-$CO_2$ chemical interaction such as $CO_2$ with superbased-derived protic ILs\cite{Wang2010}, phenolic ILs, as well as carboxylation of imidazolium acetate by $CO_2$\cite{Wang2012}. A more specific version of reactions for IL and $CO_2$ can be found in the ``Reactions for IL and $CO_2$" section of supplementary materials.

\subsubsection{Analyzing the explanation result from IL explainer}
\begin{figure}[hbt!]
    \begin{subfigure}{0.48\textwidth}
      \centering
      \includegraphics[width=\linewidth]{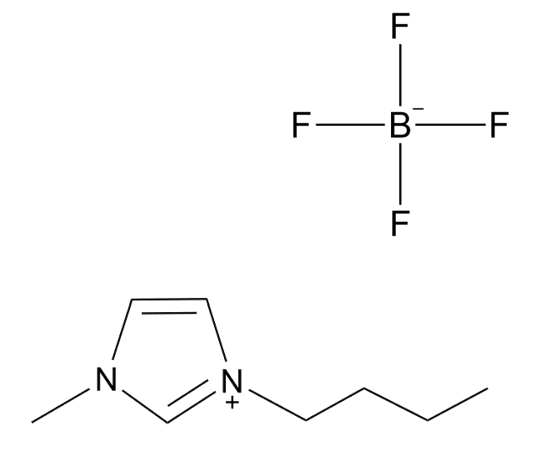}
      \caption{}
      \label{fig:ILpair}
    \end{subfigure}
    \begin{subfigure}{0.48\textwidth}
      \centering
      \includegraphics[width=\linewidth]{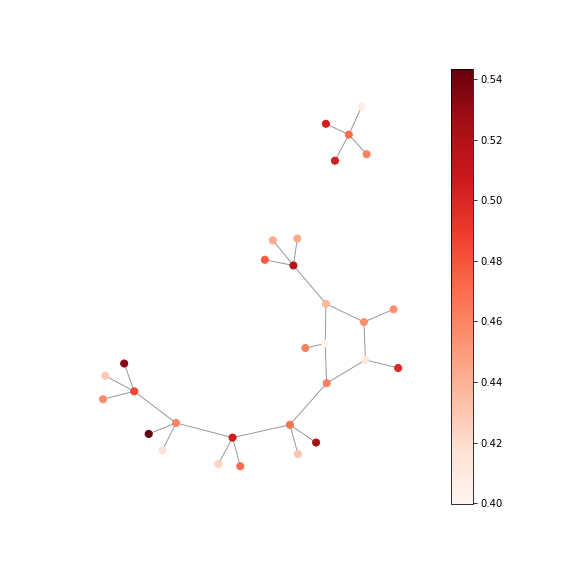}
      \caption{}
      \label{fig:hotmap}
    \end{subfigure}
    \begin{subfigure}{0.48\textwidth}
      \centering
      \includegraphics[width=\linewidth]{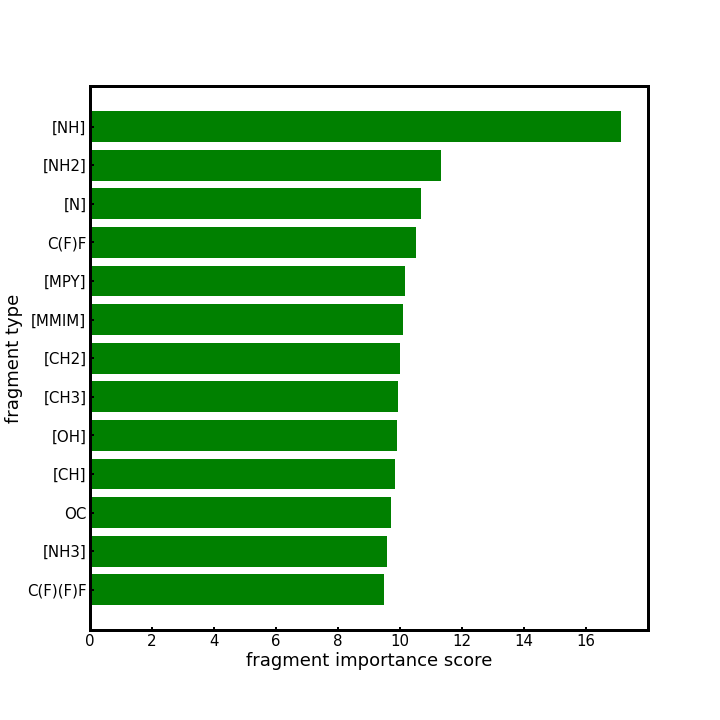}
      \caption{}
      \label{fig:ranking}
    \end{subfigure}
    \begin{subfigure}{0.48\textwidth}
      \centering
      \includegraphics[width=\linewidth]{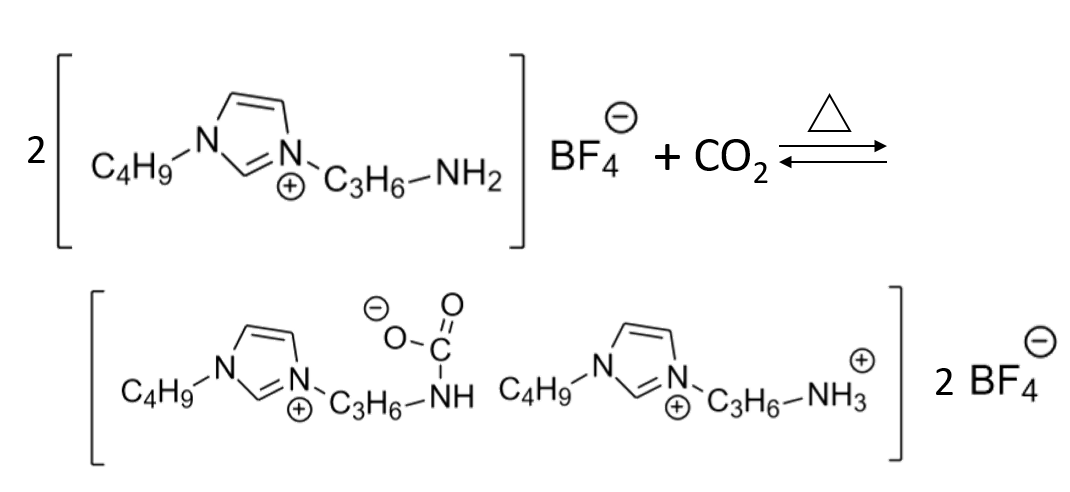}
      \caption{}
      \label{fig:NHmech}
    \end{subfigure}
\caption{(a) structure for the example molecule, here we use [$PMIM$] and [$BF_4$],(b) the importance hot map for [$BMIM$] and [$BF_4$], (c) fragment importance of cations with respect to the whole data set,(d) proposed reaction of $CO_2$ with [$NH_2p-bim$][$BF_4$]}
\label{explainres}
\end{figure}

Fig.~\ref{explainres} shows the importance analysis result from IL Explainer. Fig.~\ref{fig:ILpair} shows an example of the molecular structure of an IL molecule pair [$PMIM$] (cation) and [$BF_4$] (anion). this IL molecule pair, we use IL Explainer to get a molecule graph with node importance as shown in Fig.~\ref{fig:hotmap}. Each node (atom) importance is shown through a hot map, the importance value of each node is mapped to the color deepness. As for dataset level explanation, with the importance score function discussed in the IL Explainer section, we compute the importance score of different fragment with respect to the dataset and rank them according to the score. Due to the reason that the importance of anion and cation usually has differences according to the experiment record,\cite{Anthony2005} we only summarize fragments that appeared in cations to make the result more comparable. Finally, we have an importance ranking of fragments exist in cations with respect to the whole dataset in Fig.~\ref{fig:ranking}

In the following part, the result will be discussed case by case. First, let's take a look at fragments with medium and low ranks in the result Fig.~\ref{fig:ranking}. Based on the physical absorption theories, the type of interaction between most of the fragments in this region and $CO_2$ belong to physical interaction. Specifically, $[CH_2]$, $[CH_3]$, $[CH]$ are components of the alkyl chain, their contribution is creating more free volume for $CO_2$ to solve based on the free volume theory\cite{Huang2005,Lei2014}. Fragments like $C(F)F$, $C(F)(F)F$, $OC$ are also favorable for improving the $CO_2$ solubility with physical interaction\cite{Zeng2017}. $[OH]$ can form a hydrogen bond with $CO_2$ to strengthen their interaction. However, physical interactions are less strong than chemical interactions between ILs and $CO_2$. For this reason, it is hard for the fragment that forms physical interaction with ILs to reach a high rank among all fragments. To take a step further, we can see most of the fragments with a high ranking tends to form a chemical interaction with the $CO_2$. One of the most interesting results is the amine group, amine group is said to have a strong chemical interaction with the $CO_2$. Usually, when amine reacts with $CO_2$, one of the $N-H$ bonds of Nitrogen will be broken and the hydrogen can be substituted with $O-C=O$ (refer to mechanism in Fig.~\ref{fig:NHmech}). However, in our result, $[NH]$ and $[NH_2]$ take up the top 2 rankings, but $[NH_3]$ receives a very low ranking. To further explore the reason behind it, we combine the data points related to amine groups and the reaction condition to find the reason.

\begin{figure}[hbt!]
    \begin{subfigure}{0.47\textwidth}
      \centering
      \includegraphics[width=\linewidth]{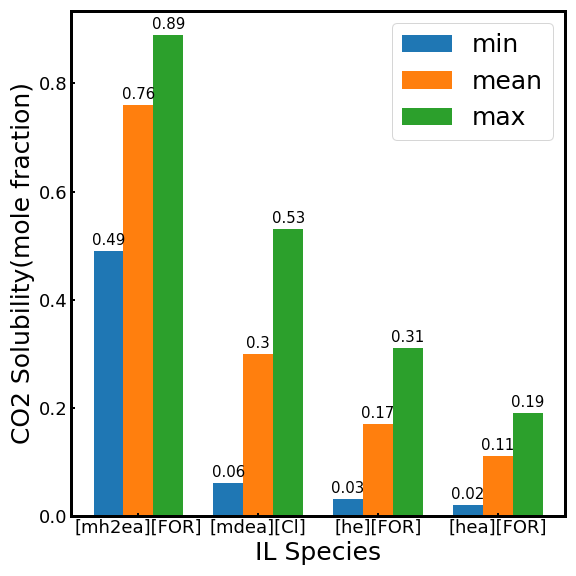}
      \caption{}
      \label{fig:Sol}
    \end{subfigure}
    \begin{subfigure}{0.47\textwidth}
      \centering
      \includegraphics[width=\linewidth]{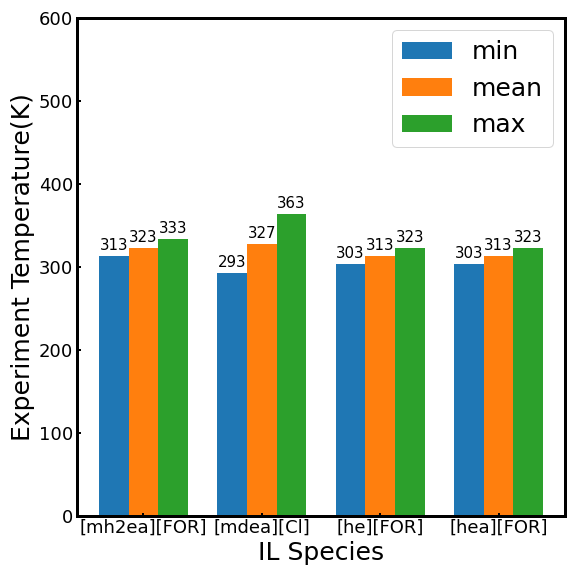}
      \caption{}
      \label{fig:Temp}
    \end{subfigure}
    \begin{subfigure}{0.47\textwidth}
      \centering
      \includegraphics[width=\linewidth]{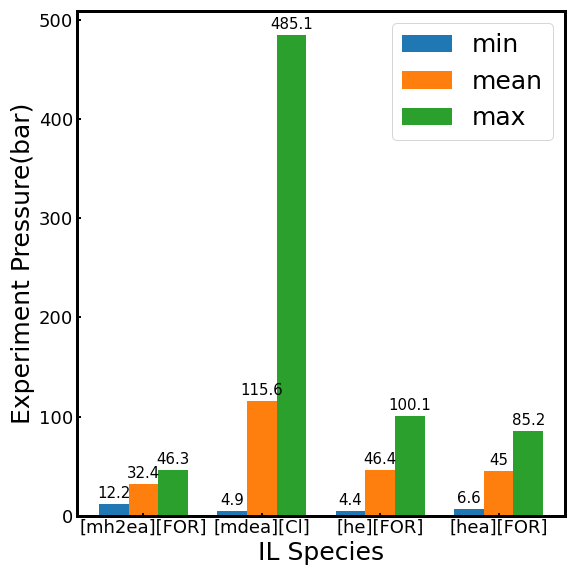}
      \caption{}
      \label{fig:Pressure}
    \end{subfigure}
    \begin{subfigure}{0.42\textwidth}
      \centering
      \includegraphics[width=\linewidth]{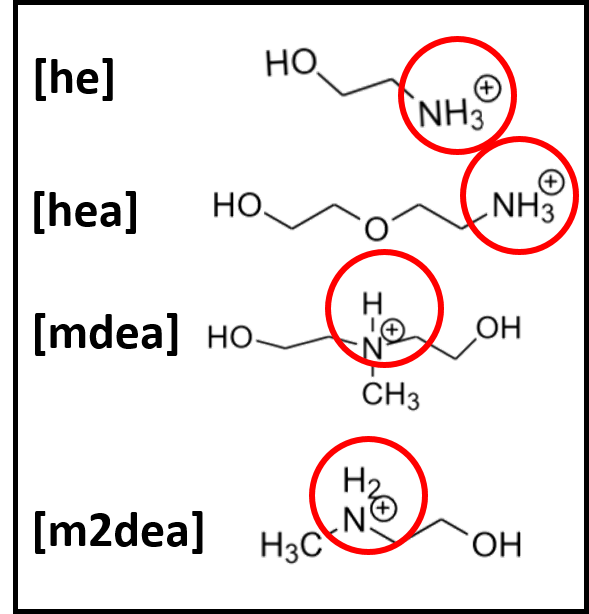}
      \caption{}
      \label{fig:NH}
    \end{subfigure}
\caption{(a) summary for solubility of amine ILs,(b) summary for the temperature of amine ILs,(c) summary for the pressure of amine ILs,(d) molecule structure of amine ILs mentioned in the figure}
\label{fig:datasetdigging}
\end{figure}

Fig.~\ref{fig:datasetdigging} contains details of the data points related to the amine group, there are four kinds of IL pairs that contain amine group $[m2hea][FOR]$ (78 data points), $[mdea][CL]$ (35 data points), $[he][FOR]$ (27 data points), $[hea][FOR]$ (21 data points). Fig.~\ref{fig:Sol}~\ref{fig:Temp}~\ref{fig:Pressure} include the max, mean, min of $CO_2$ solubility, temperature, and pressure of data points of four kinds of IL molecule pair. Fig.~\ref{fig:NH} shows the type of amine group each cation contains, including $[NH_3]$ in $[hea]$ and $[he]$, $[NH]$ in $[mdea]$, $[NH_2]$ in $[mh2ea]$. $[hea]$, $[he]$, and $[mh2ea]$ share the same anion, and the mean temperature and pressure are close. Under those conditions, $[mh2ea]$ shows a higher $CO_2$ solubility than the other two in a large amount. Considering the major structural difference of these three cations is the amine group, IL Explainer successfully captures the information where $[NH_2]$ importance is much higher than $[NH_3]$. Besides, although $[mdea]$ is paired with $[Cl]$ that is different from $[FOR]$ in the other three IL pairs, it still shows a higher $CO_2$ solution than other $NH_3$ based IL pairs under similar average pressure and temperature condition. Since $[mdea]$ contain $[NH]$ instead of $[NH_3]$, we can see that IL Explainer also detects that $[NH]$ bears higher importance than $[NH_3]$. Furthermore, if we think about this problem from an energy perspective, there is still a more fundamental reason behind it. Fig.~\ref{fig:NHmech} shows a possible reaction for $[NH_2]$ and $CO_2$. In this reaction, $[NH_2]$ plays as a reactant while its product contains $[NH_3]$. This reaction takes place with a heating condition, which means the product is more stable than the reactant in energy. In this way, we can understand $[NH_3]$ as a more stable structure than $[NH_2]$, and thus, it is harder for $[NH_3]$ to react with $CO_2$ when compared with other amine groups with less hydrogen. The most exciting thing here, however, is that IL Explainer captures the fact that $[NH_2]$ and $[NH_3]$ are more important than $[NH_3]$ in $CO_2$ absorption for structure stability reasons from a purely data-driven manner. This demonstrates the power of IL Explainer in providing insight into a chemical problem from a data perspective and the potential on assisting understanding and designing new functional IL molecules in the future. Besides, IL Explainer is highly scalable. Since the importance score for each atom within the molecule is accessible, we can gain the importance of every substructure in the molecule. For example, by collecting the importance score of the alkyl chain of different lengths, we can study how alkyl chain length affects $CO_2$ solubility in IL. Moreover, IL Explainer can be extended to periodic systems by considering the periodic interaction of nodes in the adjacency matrix as well. By this means, nodes that are neighboring under the periodic boundary condition are modeled in the interpretable GNN.

\section{Conclusion}
To summarize, we develop two categories of ML methods (descriptor-based ML models and GNNs) for predicting $CO_2$ solubility in ILs and an explanation method to detect the importance of the functional groups in a data-driven manner. For descriptor-based ML models, the result shows FP outperforms GC in most model-descriptor combinations. Among all of the experiments in this part, MLP with FP reaches the best performance of MAE (0.0151) and $R^2$ (0.9883). For the GNN model, we develop GCN, GAT, and GIN models with a virtual global node. GIN reaches the best performance with MAE (0.0137) and $R^2$ (0.9884), which demonstrates the effectiveness of GNNs in learning from graphical data. Furthermore, for the explanation method, the IL explainer takes a trained GNN and an IL molecule data point as input. By learning a mask to maximize the mutual information change, we can gain a node level importance explanation. Through statistical counting and normalization, we make a fragment importance rank for cations across the whole dataset. The ranking result shows that fragments that have physical interaction with $CO_2$ tend to have less importance than those that have chemical interaction. This can be explained as the chemical interaction is usually more strengthen compare to physical interaction. Besides that, for chemical interaction fragments, we find that amine groups with different numbers of hydrogen can be differently favorable for the absorption process. Results have shown that the amine group with less hydrogen connected to nitrogen could be more favorable in formalizing stable chemical interaction with $CO_2$. The accurate ML models and importance ranking obtained from explainable GNN can provide insights into designing new functional ILs in the future.

%% classes.  This should be given within the "acknowledgment"
\section{Code Availability}
Code and data can be found through the following GitHub link:
\href{https://github.com/ftyuejian/Predicting-CO2-Absorption-in-Ionic-Liquid-with-Molecular-Descriptors-and-Explainable-GNN}{Predicting-CO2-Absorption-in-Ionic-Liquid-with-Molecular-Descriptors-and-Explainable-GNN}. 

\begin{acknowledgement}
We thank the start-up fund provided by the Department of Mechanical Engineering at Carnegie Mellon University. The work is also funded in part by the Advanced Research Projects Agency-Energy (ARPA-E), US Department of Energy, under award no. DE-AR0001221.
% Please use ``The authors thank \ldots'' rather than ``The
% authors would like to thank \ldots''.

% The author thanks Mats Dahlgren for version one of \textsf{achemso},
% and Donald Arseneau for the code taken from \textsf{cite} to move
% citations after punctuation. Many users have provided feedback on the
% class, which is reflected in all of the different demonstrations
% shown in this document.

\end{acknowledgement}

% \bibliography{total}
\begin{suppinfo}
The Supporting Information is available. Experiment details of Descriptor-based machine learning, GNN models, Prediction-label plots, and the reaction condition and mechanism mentioned in the paper are included in the supplementary.
% This will usually read something like: ``Experimental procedures and
% characterization data for all new compounds. The class will
% automatically add a sentence pointing to the information online:

\end{suppinfo}


\bibliography{total}

\end{sloppypar}
\end{document}